\documentstyle[12pt]{article}

\begin{document}
\newcommand {\be}{\begin{equation}}
\newcommand {\ee}{\end{equation}}
\newcommand {\bea}{\begin{array}}
\newcommand {\cl}{\centerline}
\newcommand {\eea}{\end{array}}
\renewcommand {\theequation}{\thesection.\arabic{equation}}
\renewcommand {\thefootnote}{\fnsymbol{footnote}}
\newcommand {\newsection}{\setcounter{equation}{0}\section}

\def\nc{noncommutative }
\def\com{commutative }
\def\ncy{noncommutativity }
\def \simlt{\stackrel{<}{{}_\sim}}
\def \simgt{\stackrel{>}{{}_\sim}}
\baselineskip 0.65 cm
\begin{flushright}
IC/2000/03 \\
hep-th/0001089
\end{flushright}
\begin{center}
{\Large{\bf Noncommutative Super Yang-Mills Theories 

with 8 Supercharges and Brane Configurations}}

\vskip .5cm
 M.M. Sheikh-Jabbari
\footnote{ E-mail:jabbari@ictp.trieste.it } \\

\vskip .5cm

 {\it The Abdus Salam International Center for Theoretical Physics, 

Strada Costiera 11,Trieste, Italy}\\
\end{center}

\vskip 2cm
\begin{abstract}
In this paper we consider $D=4$ NCSYM theories with 8 supercharges. We study
these theories through a proper type IIA (and M-theory) brane configuration. 
We find the one loop beta function of these theories and show that there is an
elliptic curve describing the moduli space of the theory, which is in
principle the same as the curve for the commutative counter-part of our theory. 
We study some other details of the dynamics by means of this brane
configuration.
\end{abstract}
\newpage
\section{Introduction}

Recently it has been shown that the noncommutative gauge theories can arise as the 
low energy effective open string theory in the presence of D-branes with non-zero
NSNS two form, B-field, on them \cite{{HD},{CK},{Sh1},{HV},{SWN}}.

The action for the \nc gauge theories is obtained from the usual (\com )
Yang-Mills theories by replacing the ordinary product of the fields by the 
*-product (for a review see \cite{SWN}). From now on we call them NCYM theories.
It has been shown that the \nc $U(1)$ theory can be understood as the \com gauge
theory with a deformed gauge group, so that the dynamics of this new gauge theory is 
governed by a series of dipoles in addition to the usual photons \cite{{Sh2}}.
Similar statements can be made for NCYM with any gauge group. It has been
discussed that NCYM theories are renormalizable \cite{{Sh2},{MS},{KW}}.
Besides the pure gauge theories, the \nc scalar field theories have also been
considered and it has been argued that these \nc gauge theories are renormalizable
if and only if their \com limit (\nc parameter, $\theta$, equal to zero) is
renormalizable \cite{{chep},{Shir}}.   
  
From the string theory side, the branes with constant B-field on them preserve 16
supercharges, half of the supercharges of the type II theories, similar to the
usual D-branes, and hence when we study the low energy effective theory of such
branes we actually deal with a NC (Super)YM (NCSYM) theory with 16 supercharges.
Although the perturbative fermionic fields in a \nc background have not been well
elaborated, since the \nc field theories can be understood as a \com field
theory with definite interaction terms, it is believed that these theories should
have the usual feature of the gauge theories with 16 supercharges. As an example, it
has been shown that the large N limit of the NCSYM can be studied through the gauge
theory/gravity on definite background correspondence \cite{{MR},{HI},{AOS}}. 
The conceptual new point of these NCSYM theories is that, despite being maximally
supersymmetric, unlike the usual SYM (with 16 supercharges), they are {\it not
conformal invariant} and this is because the deformation parameter, $\theta$, is a
dimensionful parameter with dimensions of (length)$^2$.

For the ${\cal N}=4,D=4$ SYM theories large amount of supersymmetries kill all the
interesting dynamics. In this paper we try to study more interesting theories
with less supersymmetries, namely  ${\cal N}=2,D=4$ NCSYM. After the work of 
\cite{SW} we learnt how we can study the four dimensional ${\cal N}=2$ SYM theories,
their BPS spectrum, their moduli space of vacua and its singularities, and how we
can extract all of these physics from a {\it proper elliptic curve}.
In another paper, \cite{Witt}, it was shown how the curve itself, and
hence all of the above mentioned physics of ${\cal N}=2,D=4$ SYM theories, can be
obtained from the definite type IIA string and M theories {\it brane configuration}
\footnote{In that paper only the $\prod_{\alpha=1}^{N} SU(k_{\alpha}$) gauge group
was
considered, later the same procedure generalized to SP and SO gauge theories in
extensive papers, which we are not going to list them here.}.

In this paper, we show how the brane configuration method should be extended
and modified for studying the ${\cal N}=2,\ D=4$ NCSYM with $SU(N)$ gauge group.
In this way we study some of the physical aspects of the deformed SYM theories, and 
discuss that most of the physics expected in the \com ${\cal N}=2,\ D=4$ SYM
hold in the \nc case too. Intuitively this should be related to the fact that 
the number of degrees of freedom of \nc gauge theories, at least for the planar
diagrams, are the same as the \com gauge theories \cite{{BS},{Sh2}}.

The paper is organized as follows. In the next section we fix our conventions and
notations and review the preliminaries we need.
In section 3, we build the proper brane configuration from type IIA NS5-brane and
(D4-D2)-brane bound states so that the brane system preserves 8 supercharges. In
section 4 we relate the (pure) $D=4$ NCSYM to the brane configuration of the
section 3 and identify the parameters of gauge theory in terms of the brane
model parameters. In this way we find the one loop beta function of the theory. By
lifting the brane configuration to M-theory we study the Coulomb branch of the 
theory. Moreover we discuss different phases of our \nc theory in different energies. 
The last section is devoted to concluding remarks and open questions.

\section{ Conventions and Preliminaries}
\setcounter{equation}{0}
{\it a)String Theory} 

It has been shown that turning on B-fields polarized parallel to a $D_p$-brane,
is equivalent to building various bound states of $p,\;p-2,\;p-4,...$ branes,
depending
on the rank of the B-field \cite{{More}}. For the B-field of rank one, 
these bound states are composed of a p-brane and a uniform distribution of
(p-2)-branes on it. So that if the worldvolume of the $p$-brane is located along
$012...p$ directions and $B_{p-1,p}$ is the non-zero component, then the
(p-2)-branes span $012...(p-2)$ directions. This argument can be simply generalized
to any higher rank B-fields. In our notations $B$ is a dimensionless parameter and
$l_s^{-2}B$ is the (p-2)-branes density. The mass density of the bound state
in this notation is $(l_s^{(p+1)}g_s)^{-1}\sqrt{1+B^2}$ \cite{More}.    

The perturbative dynamics of these bound states, similar to the individual D-branes,
is governed by the open strings ending on them, but this time because of the
B-field, they should satisfy mixed boundary conditions. Studying such open strings
we learn that the brane worldvolume is in fact a \nc plane:
\be\label{theta}
[x^{\mu},x^{\nu}]=il_s^{2}({B \over 1-B^2})^{\mu\nu}\equiv i\theta^{\mu\nu}.
\ee
As we see $\theta$ is a parameter with dimensions of (length)$^2$.
\vskip .5cm
{\it b)Noncommutative Field Theories}
\vskip .5cm

Field theories on a \nc plane are obtained from their commutative version by
replacing the product of the fields by the *-product defined as follows\cite{SWN}

\be
(f*g)(x)=exp({i\over 2}\theta_{\mu\nu}\partial_{x_{\mu}}\partial_{y_{\nu}}) 
f(x)g(y)|_{x=y}.
\ee
Besides the field products, to couple the \nc theories to a NCYM, we should
replace the ordinary derivative with a "covariant derivative":

\be
\partial_{\mu} * \rightarrow \partial_{\mu} *\; +\{A_{\mu},\;*\}_{M.B.},
\ee
where the "Moyal Bracket" in the above relation is $\{f,g\}_{M.B.}=f*g-g*f$.
As we see for the slowly varying fields ($|{\partial f\over f}|^2<<|\theta |^{-1}$),
these \nc field theories effectively behave like the \com theories. 

It is worth noting that for the case of \nc SU(N) gauge theory, when $A_{\mu}$ take
values in SU(N) algebra, the field strength $F_{\mu\nu}$ defined by (2.3) is not algebra
valued unlike the usual gauge theories (it is sitting in the SU(N) group). 
   
\section{The Model with IIA branes}
\setcounter{equation}{0}
In order to reduce the number of supersymmetries, as it has been shown and discussed
in \cite{{Witt},{HW}}, we consider the brane intersections.

Since we are interested in the four dimensional theories, like \cite{Witt}, we
consider the IIA theory and its NS5- and (D4-D2)- branes.
Let us consider the following brane configuration:

(N+1) NS5-branes labelled by $\alpha=0,1,...,N$, spanning $012345$ directions.
These fivebranes are located at different values of $x^6$ direction, and 

a number of D4-branes with their worldvolume along $01236$ so that there are
$k_{\alpha}$ number of them between the $(\alpha-1)^{th}$ and $\alpha^{th}$
NS5-branes. As we see the intersection of these branes is a 3+1 dimensional space,
$0123$, on which we have our ${\cal N}=2$ gauge theory. 

To find a NCSYM, we turn on B-field along the D4-branes. We should recall that
turning on a B-field will not change the number of conserved supercharges (see the 
Appendix for some explicit calculations).
From the superalgebra point of view, turning on the B-field corresponds to choosing
another central extension of the algebra (different from that of the individual
brane). In other words, when we turn on the B-field, we again find a BPS solution
which preserves {\it half} of supersymmetries.

There are various choices for the B-field, it can be of rank one or two, and also 
it can have different polarizations. In this paper we present calculations for
the rank one case, with non-zero $B_{23}=B$, which is the most interesting case. We
will discuss the $B_{36}$ and rank two cases briefly in the discussion section.
With a non-zero $B_{23}$, our brane configuration consists of NS5-branes along
$012345$ and D4-branes along $01236$ and a distribution of D2-branes having
their worldvolume along $016$.
These D2-branes are {open} D2-branes as discussed in \cite{Stro}.

In order to find a system with finite energy we should compactify the $x^6$
direction, or equivalently we should identify the $(N+\alpha)^{th}$ fivebrane with
$\alpha^{th}$. 

Along the lines of \cite{Witt}, the $x^6$ coordinate as a function of
$v\equiv x^4+ix^5$, is obtained by minimizing the total fivebrane worldvolume. For
the large $v$ equation of $x^6$ reduces to a source free Laplace equation,
\be
\nabla^2 x^6=0.
\ee
Since $x^6$ is only a function of the directions normal to the brane bound state,
\be
x^6={\cal K} ln|v|+constant.
\ee
The parameter ${\cal K}$ is actually the ratio of the tensions (or mass densities)
of two intersecting branes:
\be\label{calK}
{\cal K}\sim {{\rm (D4-D2)\; mass\; density}\over {\rm NS5-brane\; mass\; density}}
\sim{l_s^{-5}g_s^{-1}\sqrt{1+B^2}\over l_s^{-6}g_s^{-2}}=l_s g_s\sqrt{1+B^2}.
\ee

For the case with $k$ number of these bound states on top of each other, obviously
our ${\cal K}$ factor should be multiplied by a factor of $k$.
It is worth noting that (\ref{calK}) is the $B=0$ result with $g_s$ replaced
with the open string coupling \cite{SWN} and, it reduces
to the results of \cite{Witt} for the $B=0$.
Since the (D4-D2)-branes ending on the fivebrane on its left pulls it in the
opposite direction compared to those ending on its right, the ${\cal K}$ factor for
them
should have different relative sign. If we have $q_L$ (and $q_R$) (D4-D2)-branes
ending on the fivebrane on its left (and right), which are located at 
$a_i, i=1,....,q_L$ (and $b_j, j=1,....,q_R$), then the asymptotic form of $x^6$ is
\be\label{x6}
x^6={\cal K}(\sum_{i=1}^{q_L} ln|v-a_i|- \sum_{j=1}^{q_R} ln|v-b_j|)+constant.
\ee
A well-defined $x^6$ for $v\rightarrow \infty$ is obtained if and only if $q_L=q_R$.

To study the gauge theory dynamics we also need to consider the moving
(D4-D2)-branes. This motion is realized by letting $a_i$ and $b_j$ become
functions of
the space-time coordinates ($0123$). This motion as explained in \cite{Witt},
corresponds to the question of IR divergences of the NCSYM\footnote{As we will
discuss in the next chapter, the IR behaviour of the NCSYM is the same as the \com 
SYM.}. This motion contributes to the fivebrane kinetic energy as
$\int d^4xd^2v \partial_{\mu}x^6\partial^{\mu}x^6,\;\mu=0,1,2,3$. We should note
that the metric on our space-time is the {\it open string metric}. With $x^6$ given
by (\ref{x6}), it becomes    
\be
\int d^4xd^2v |{\rm Re}\biggl((\sum_i\partial_{\mu}a_i({1\over v-a_i})-
\sum_j\partial_{\mu}b_j({1\over v-b_j})\biggr)|^2.
\ee
The integral converges if and only if
\be
\sum_i a_i-\sum_j b_j=q_{\alpha},
\ee
where $q_{\alpha}$ is a constant, and is a characteristic of the $\alpha^{th}$
fivebrane.
The $q_{\alpha}$'s are determined by the separation between (D4-D2) branes, and hence
we expect them to be related to the "bare mass" of the gauge theory hypermultiplets.

The quantum mechanical treatment of our \nc gauge theory corresponds 
to the M-theory limit of our brane model. In
that limit type IIA on ${\bf R}^{10}$ is actually the M-theory on ${\bf
R}^{10}\times S^1$, with radius $R=l_sg_s$. But, this is only the compactification
radius for closed strings, i.e. $R$ is the radius which relates the 10 and 11
dimensional supergravities. As we discuss in the following, the open string
compactification radius is different\footnote{The fact that open string
compactification radius, $R_o$, is effectively different, in the work of 
Connes, Douglas and Schwarz \cite{CDS}, which was the light-like compactification
of M-theory with C-field background, is reflected in their dim${\cal H}$={\bf
Tr1}(the dimension of Schwartz space), and in our case is related to ${R_o\over
R}$.}.

In the M-theory limit, our (D4-D2)-bound state is lifted to a (M5-M2)-bound state.
Such bound states as discussed in \cite{{ST},{SWN}} are formed by turning on a
non-zero C-field background of 11 dimensional theory on the M5-brane.
This corresponds to turning on a self-dual three form of the
six-dimensional M5-branes worldvolume theory. In other words, in the M-theory limit
$B_{23}$ will be replaced by $C_{23(10)}$ and $C_{016}$, and these fields are equal 
because of the self-duality condition. From the 10 dimensional point of view, this
self-duality is related to the fact that, D2-branes are the sources for the $RR$
three form of type IIA theory, and in our case their density is given by $B_{23}$.
Having this in mind, we learn that the actual compactification radius viewed from
the fivebrane worldvolume theory, and hence from the open M2-branes point of view, 
is not $R$, but $R\sqrt{1+C^2}$. Since the eleven dimensional C-field 
and the ten dimensional B-field are supergravity fields they should be
related by closed string parameters, which in our conventions are
\be
l_p^{-3}R C=l_s^{-2}B.
\ee
So, the effective compactification radius for the open strings, is 
\be
R_o=R\sqrt{1+B^2}.
\ee 
 
If we denote the eleventh dimension by
$x^{10}$, which is periodic with periods $2\pi R$, we have
\be
x^6+ix^{10}=R \sqrt{1+B^2}\big\{(\sum_{i=1}^{q_L} ln|v-a_i|-
\sum_{j=1}^{q_R} ln|v-b_j|)+
i\phi\big\},
\ee
where $0\leq\phi={x^{10}\over R_o}<2\pi$, and hence

\be\label{sv}
s\equiv {x^6+ix^{10}\over R}=\sqrt{1+B^2}\{\sum_{i=1}^{q_L}
ln(v-a_i)-\sum_{j=1}^{q_R}ln(v-b_j)\}+constant.
\ee

The fact that $s$ is {\it holomorphic} in $v$, is expected from the supersymmetry. 
Let us  concentrate on the imaginary part of (\ref{sv}). It states that
circling around one of $a_i$ and $b_j$'s 
in the $v$ complex plane, $x^{10}$ jumps by $\pm 2\pi R_o$. From the fivebrane
worldvolume effective theory, the end points of (D4-D2)-brane bound states are
viewed as "dyonic vortices". In the T-dual version, where our brane configuration is
composed of IIB NS5-branes (spanning $012345$) and (D3-D1) bound state (with
D3-branes along $0126$ and D-strings along $06$), the end point of the (D3-D1)-brane
looks like "dyons" of fivebrane effective theory. With our conventions in definition
of ${\cal K}$, (\ref{calK}), these dyons carry one unit of magnetic charge and 
$l_s^{-2}B V_2$ units of electric charge ($V_2$ is the volume of $12$ plane).

\section{ Four Dimensional Noncommutative Gauge Theory}
\setcounter{equation}{0}
Now let us return to our main question: what can we learn about the $D=4$ NCSYM from
the above brane configuration.

Since we have considered the $B_{23}\neq 0$, according to (\ref{theta}) we deal with 
a NCSYM with non-zero $\theta_{23}$,
\be
\theta\equiv\theta_{23}=l_s^2{B\over 1+B^2}.
\ee

If we consider the brane configuration we built in the previous section, because of 
(\ref{x6}), we deal with $\prod_{\alpha=1}^N\; SU(k_{\alpha})$ NCSYM theory.

Now we should identify our gauge theory coupling constant through the string theory
parameters.
The naive answer to this question can be understood in light of the \com /\nc
gauge theory correspondence proposed in \cite{SWN}, by expanding the Born-Infeld
action for the D4-branes with non-zero B-field, up to the first order in $\alpha'$.
Then the gauge coupling $g_{\alpha}$ of the $SU(k_{\alpha})$ should be given by
\be\label{g2}
{1\over g^2_{\alpha}(v)}={x^6_{\alpha}(v)-x^6_{\alpha-1}(v)\over l_sg_s}.
\ee
Replacing $x^6$ from (\ref{x6}) we find that the above relation is exactly the
same as its \com counter-part, with $g_s$ which is the closed string
coupling, replaced with {open string coupling}.

According to (\ref{g2}) we propose that $g_{\alpha}^{-2}$ for the NCSYM to be
logarithmic divergent for large $v$. This is the familiar asymptotic behaviour
of the \com theory at high energies. This proposal is expected since we believe that
the \nc gauge theories can be mapped into a \com gauge theory through the Fourier
expansion \cite{{Sh2},{Shir}} and also, we know that the planar degrees of
freedom of the NCSYM is the same as its \com counter-part \cite{BS}.  
Moreover we propose that this logarithmic divergence corresponds to the one loop
beta function of our four dimensional NCSYM. This is in exact agreement with
perturbative results of \cite{{Sh2},{Shir},{BS}}.

The theta angle of the $SU(k_{\alpha})$ theory, $\theta_{\alpha}$, is then naturally
related to the imaginary part of (\ref{sv}) as\footnote{$\theta_{\alpha}$ should
not be confused with $\theta$, the \ncy parameter.}    

\be
\theta_{\alpha}={x^{10}_{\alpha}-x^{10}_{\alpha-1}\over R}.
\ee
Hence the 
\be
\tau_{\alpha}(v)={\theta_{\alpha}\over 2\pi}+{4\pi i\over g^2_{\alpha}}=
i(s_{\alpha}(v)-s_{\alpha-1}(v)).
\ee

In the large $v$ limit:

\be\label{tau}
\tau_{\alpha}(v)=i\sqrt{1+B^2}(2k_{\alpha}-k_{\alpha-1}+k_{\alpha+1})ln v.
\ee
As we see the only difference of (\ref{tau}) with the \com version is the
coefficient in front. 

Comparing the standard one loop asymptotic freedom formula, $\tau=ib_0ln v$, we find
the important result 
that the one loop beta function for NCSYM theories with the above mentioned
gauge group is proportional to the one loop beta function of the  \com counter-part. 

Analogous to the \com case, the open strings having their end points on the open 
(D4-D2)-bound states on the opposite sides of fivebranes form the {\it
hypermultiplets} of our NCSYM theory, obviously in the $(k_1,{\bar k}_2)\oplus
(k_2,{\bar k}_3)\oplus ....(k_{N-1},{\bar k}_N)$ representation. 
The only point which one should note here is that the mass of these hypermultiplets
should be calculated by squaring the momentum with the open string metric
\cite{SWN}. These hypermultiplets become massless classically when
(D4-D2)-bound states on the right of a fivebrane end on the same point as the 
(D4-D2)-branes on the left. Like the \com case, when one these hypermultiplets 
becomes massless, namely $(k_i,{\bar k}_{i+1})$, for the $k_i=k_{i+1}$  we
expect the related factor of the gauge group to become a $SU(k_i)$ instead of 
$SU(k_i)\times SU(k_{i+1})$. 

Up to here we have identified and related the parameters of NCSYM with those of our
brane configuration model and briefly discussed the similarities of \nc and
\com theories at one loop. Now we want to study different phases of 
the \nc gauge theories.

As we mentioned the \ncy parameter, $\theta$, is a dimensionful parameter and hence
there are two parameters of energy dimensions, $\theta^{-1/2}$ and $R_o^{-1}$ ( the
"open string" eleven dimensional radius), with which we should compare our energy
scales. We distinguish two different phases:

{\it a) ${\theta/R_o^2}\simlt 1$}   
\vskip .25cm

In this case our \nc theory effectively behaves as a \com SYM because, before the
energies that \ncy effects show themselves, $\theta E^2\sim 1$, we should uplift 
our theory to M-theory, where we only have (M5-M2)-bound state effective theory
which is not a \nc one.

If we replace the $\theta$ and $R_o$ by their string theoretic values we have  
\be\label{Rthet}
{\theta \over R_o^2}\sim {l_s^2\over l_s^2g_s^2}{B\over (1+B^2)^2}.
\ee
Since ${B\over1+B^2}\leq 1/2$,  for ${\theta/R_o^2}\simlt 1$ cases we have
\be\label{gos>}
g_{{\rm open\;string}}\simgt 1,
\ee 
and hence this case for generic B, corresponds to the strong coupling of our \nc
gauge theory (which as we argued above) is not a \nc theory. For very large
B, (\ref{gos>}) reads as $ g_{YM}B\simgt 1$ which
corresponds to the \com gauge theory weak coupling.

{\it b) ${\theta/R_o^2}\simgt 1$}   
\vskip .5cm

This is the more interesting case. From (\ref{Rthet}) we see that for generic B, 
we are actually dealing with weakly coupled gauge theory, and the theory shows
different behaviours at different energies:

{\it b-1) IR limit (${\theta E^2}<< 1$)}:
\vskip .25cm
In this regime the \nc effects can totally be neglected and we effectively see the
\com SYM theory. So, all the arguments of IR limit of \com SYM holds for the
NCSYM too.

{\it b-2) Intermediate energies (${1\over\sqrt{\theta}}\simlt E<< {1\over R}$)}:
\vskip .25cm
This is the phase which we actually deal with a NCSYM theory.

{\it b-3) UV limit (${E}\simgt {1\over R_o}$)}:
\vskip .25cm
In this case we should uplift our brane configuration to the M-theory, where there
is no \nc description. 
\vskip .5cm
{\it M-Theory Interpretation and Seiberg-Witten Curves}
\vskip .5cm
In order to study the quantum structure of NCSYM, we need to lift the brane
configuration we
built in IIA theory (which describes the classical regime) to the M-theory.
In the M-theory, as we stated earlier the (D4-D2)-bound state coincides with the
(M5-M2)-bound state in the M-theory and NS5-branes are viewed as M5-branes.
The (M5-M2)-brane, is actually a fivebrane with a non-zero self-dual three form on
it. Hence, in the M-theory limit our IIA brane configuration is reinterpreted as
a single fivebrane with a definite C-field on it, whose worldvolume sweeps
$R^4\times \Sigma$ where $\Sigma$ is a hyper-surface given by a {\it holomorphic}
function in the $v$ and $s$ complex planes. Since $s$ is not single valued, it is
more convenient to introduce the $t=exp(-s/\sqrt{1+B^2})$ instead. Analogue of the
Seiberg-Witten elliptic curve, is exactly the curve defining $\Sigma$ in the 
$t$ and $v$ plane, namely, $F(t,v)=0$. Compared to the \com case the only difference 
is in the definition of $t$ with respect to $s$.
However, the powers in $v$ will remain the same as the \com case.

One can study the low energy four dimensional \nc theory  from the six dimensional
effective theory of M5-branes, if we denote the 
field strength of the two-form field living on the M5-brane by $T$, we know that
$T$ is self-dual and its equation of motion is $dT=0$. One can decompose
$T$ as follows:
\be
T=F\wedge \Lambda+ ^*F\wedge ^*\Lambda+ C,
\ee
where $C$ is the self-dual constant background and in our case has non-zero
components in $016$ (and $23(10)$).  
$\Lambda$ is a harmonic one form defined on the $\Sigma$, satisfying,
$d\Lambda=d^*\Lambda=0$. 
As discussed in \cite{Witt}, to see the reasonable low energy theory we need to
define $\Lambda$ on a point-wise compactified version of $\Sigma$,  $\bar\Sigma$,
which is obtained by adding (N+1) points to the  $\Sigma$.  $\bar\Sigma$
is a surface of genus $g=\sum_{\alpha=1}^N(k_{\alpha}-1)$.
$F$ is a two form and {\it \nc} field defined on $R^4$, obeying $DF=D^*F=0$,
where $D$'s are the \nc "covariant derivative". The low energy theory for the $g$
dimensional $\Lambda$, {\it the Coulomb branch of our theory}, is (NC $U(1)$)$^g$. 
Apart from the changes mentioned above, all the other
results and discussions of \cite{Witt} are valid in our case too.     

\section{Concluding Remarks and Discussions}

In this paper we studied the ${\cal N}=2,D=4$ NCSYM theories with gauge group
$\prod_{\alpha=1}^N SU(k_{\alpha})$ through the {\it brane configuration method}.
In order to find the NCSYM gauge theory, we considered the D4-branes with a B-field
background. We should emphasize that actually the \nc gauge theories we
are discussing are formulated on a space with open string metric
\cite{SWN}. 
By means of our brane system, we found the one loop beta function of these \nc
theories to be logarithmically divergent. Moreover, we argued that one can
still find
a {\it holomorphic} curve which describes the moduli space of our theory.
For the case at hand, the pure gauge theory, we briefly discussed the 
behaviour of the theory in different energy regimes and its Coulomb branch.

In this paper we discussed the rank one B-field along $23$ directions.
For the $B_{16}$ case, since the \nc coordinates are 3 and 6 directions, our 
four dimensional space-time ($0123$ directions) is {\it not} noncommutative and
hence we have our usual \com field theory. In addition, since instead of
NS5-branes we
have (NS5-D2)-brane bound states, we expect the equation of motion for $x^6$
not to be altered compared to the \com case.
For the same reasons for the case of  rank two B-field, i.e. $B_{23},B_{16}\neq 0$,
we expect to see the physics similar to $B_{23}\neq 0$. However, the $B_{23}=B_{16}$
seems to be an interesting special case to be discussed.  

Another interesting extension of this work is adding matter fields and the Higgs
branch of NCSYM. We believe that like the \com version this can be done by adding
D6-brane to the brane configuration. We hope to come back to this question in future
works.

There are many other interesting problems which can be addressed, e.g. the
generalization of Hanany-Witten work\cite{HW}, and (2+1) dimensional NCSYM, 
the \nc two dimensional ${\cal N}=(4,4)$ theories, theories with lower 
supersymmetries (four supercharges) which are obtained by rotating one of the
NS5-branes.   

\vskip 1cm

{\bf Acknowledgements}
\vskip .5cm
I would like to thank M. Alishahiha for many fruitful discussions and careful reading of 
the manuscript. I would like to thank H. Arfaei for discussions.

This research was partly supported by the EC contract no. ERBFMRX-CT 96-0090.
\vskip .5cm
{\large\bf Appendix}  

The explicit form of the supercharges for a (D4-D2) bound state, resulting from a
non-zero $B_{23}=B$ component, can be written as a linear combination of 
the type IIA supercharges, $Q_R$ and $Q_L$ ($\Gamma^{01...9}Q_R=Q_R$
and $\Gamma^{01...9}Q_L=-Q_L$)\cite{{HW}} : 
$$
Q=\epsilon^R\; Q_R+\epsilon^L\; Q_L,
$$
where $\epsilon^R$ and  $\epsilon^L$ are 16 component spinors satisfying the
following conditions:
$$
(I)\left\{\bea{cc}
({1\over \sqrt{1+B^2}}+{B\over \sqrt{1+B^2}}\Gamma^{23}) 
\Gamma^{016}\epsilon^R=\epsilon^L \\
\Gamma^{45789}\epsilon^R=\epsilon^L
\eea\right.
$$ 
The above conditions can be obtained from the T-dual version of the rotation
matrices in $23$ plane. The $\Gamma^{2}$ is rotated by $R(\phi)\Gamma^{2}$, with
$R(\phi)=exp(\phi\Gamma^{23})=\cos\phi+\Gamma^{23}\sin\phi$
\cite{Angle}. By T-duality $\tan\phi$ is replaced by $B$. It can easily be shown
that relations (I) allow 16 supersymmetries. Moreover one can check that for the
$B=0$ and $B\rightarrow \infty$ these conditions coincide with the supersymmetries 
preserved by a D4- and D2- brane respectively. When we also consider the NS5-branes
as well as (I), we should consider 
$$
(II)\left\{\bea{cc}
\Gamma^{012345}\epsilon^R=\epsilon^R \\
\Gamma^{012345}\epsilon^L=\epsilon^L
\eea\right.
$$ 

The simultaneous solutions of (I) and (II) give the allowed 8 supercharges.

\end{document}